\documentclass[a4paper,psfig,12pt ]{article}
\usepackage{graphics}
\begin{document}
\bibliographystyle{plain}
%\runninghead{Orbital studies of 4U~1538--52}

\title{Orbital evolution and orbital phase resolved spectroscopy of the
HMXB pulsar 4U~1538--52 with RXTE-PCA and BeppoSAX}

\author{U. Mukherjee$^{1}$, H. Raichur$^{1,2}$, B. Paul$^{1}$, 
S. Naik$^{3}$, and N. Bhatt$^{4}$}

\date{}
                                                                                                                            
\maketitle
                                                                                                                            
\begin{center}

$^{1}$Tata Institute of Fundamental Research, \\
Homi Bhabha Road, Colaba, Mumbai 400 005, INDIA \\~\\

$^{2}$Joint Astronomy Program, Indian Institute of Science, \\
Bangalore-560 012, INDIA \\~\\

$^{3}$Institute of Space and Astronautical Science, \\
3-1-1 Yoshinodai, Sagamihara, \\
Kanagawa 229 8510, JAPAN \\~\\

$^{4}$Nuclear Research Laboratory, \\
Bhabha Atomic Research Centre, \\
Mumbai-400 085, INDIA \\~\\

\end{center}

\begin{abstract}

We report here results from detailed timing and spectral studies of the
high mass X-ray binary pulsar 4U~1538--52 over several binary
periods using observations made with the Rossi X-ray Timing Explorer (RXTE)
and BeppoSAX satellites. Pulse timing analysis with the 2003 RXTE data
over two binary orbits confirms an eccentric orbit of the system.
Combining the orbitial parameters determined from this observation with
the earlier measurements we did not find any evidence of orbital decay in
this X-ray binary. We have carried out orbital phase resolved spectroscopy
to measure changes in the spectral
parameters with orbital phase, particularly the absorption column density
and the iron line flux. The RXTE-PCA spectra in the 3--20 keV energy range
were fitted with a power law and a high energy cut-off alongwith a Gaussian
line at $\sim$ 6.4 keV, whereas the BeppoSAX spectra needed only a power
law and Gaussian emission line at $\sim$ 6.4 keV in the restricted energy
range of 0.3--10.0 keV. An absorption along the line of sight was included
for both the RXTE and BeppoSAX data. The variation of the free spectral
parameters over the binary orbit was investigated and we found that the
variation of the column density of absorbing material in the line of sight
with orbital phase is in reasonable agreement with a simple model of a 
spherically symmetric stellar wind from the companion star.  \\

{\bf keywords:} Stars: Binaries: Eclipsing, Stars: Pulsars: Individual: 4U 1538-52, X-Rays: Stars

\end{abstract}

\section{Introduction}

The orbits of binary systems evolve with time due to orbital angular momentum 
changes. This change of orbital angular momentum can happen due to mass loss 
from the system, mass transfer within the system and tidal interactions between
the components. X-ray binary pulsars are suitable candidates to study such 
orbital evolution because of the following reasons. Firstly, in all these 
systems the normal companion star is transferring matter to the compact star 
either through wind capture or through Roche-lobe overflow. Secondly, the strong
gravitational pull of the compact object raises tides on the companion star 
which tend to circularize and synchronize the binary orbit. This causes orbital
angular momentum changes which result in the change of orbital period. Thirdly,
pulse timing analysis allows us to obtain an orbital ephemeris history with
an accuracy upto a few seconds. Thus monitoring X-ray binary pulsars over a
time span of a few decades 
allows us to measure the rate of change of the orbital period.
For most X-ray binaries the 
orbital ephemeris clearly deviates from the linear trend. Including a quadratic
term improves the fit in several cases (Cen X-3, Nagase et al. 1992; LMC X-4, 
Levine et al. 2000, Naik \& Paul 2004; SMC X-1,Wojdowski et al. 1998) indicating
a non-vanishing orbital period derivative. The many X-ray satellite missions 
have thus made it possible to measure the rate of change of the orbital period, showing
that for High Mass X-ray Binaries (HMXB) the orbital evolution time scales are
of the order of a million years.

The X-rays emitted from the compact object in HMXBs 
and subsequently reprocessed in the stellar wind can be used to probe the 
stellar wind of the massive companion, a supergiant 
or a Be star. An investigation of the X-ray spectrum 
from the compact object at different orbital phases can give useful 
insight into the morphology of the stellar wind. The absorbing column
density in the line of sight would vary with orbital phase and is a 
function of the mass-loss rate, stellar wind velocity and acceleration method
and also depends on any anisotropy in the stellar wind structure.
Moreover, a binary system with an elliptical orbit would be  
a better candidate in this type of study since it would have a 
larger variation of the parameters that determine the column density 
at different orbital phases. If the compact object 
is a pulsar, the orbit can often be determined precisely, which is an 
added advantage. Orbital phase resolved spectroscopy was carried out
recently on some HMXB pulsars like X 1908+075 (Levine et al. 2004) and 
GX 301--2 (Mukherjee and Paul 2004) and very different results were 
obtained.
 
4U~1538--52 is an X-ray pulsar, first detected with the UHURU satellite
(Giacconi et al. 1974). Regular X-ray pulsations with
a period of 529 s were later discovered 
with Ariel 5 and OSO-8 observations  by Davison et al. (1977). 
In addition, the OSO-8 observations revealed a clear orbital 
modulation of the pulsation period and also showed good evidence
for an eclipse lasting 0.6 day recurring with a period of 3.75 days.
The optical counterpart was found 
to be an early B type supergiant star with H-alpha emission lines 
(Parkes et al. 1978). Features in the optical spectrum suggested a 
distance to the source of $\sim$ 5.5 kpc and the mass-loss rate for the 
companion was estimated to be $\sim$ 10$^{-6}$ M$_{\odot}$ yr$^{-1}$.

Makishima et al. (1987) using TENMA and Robba et al. (1992) using
EXOSAT observed pulse periods of $\sim$ 530 s with an average spin-down
rate of $\dot{P}$ $\sim$ 3.9 $\times$ 10$^{-9}$ s s$^{-1}$. 
Subsequently, the Burst and Transient Source Experiment (BATSE) 
onboard the Compton Gamma Ray Observatory (CGRO) detected a
reversal of the long term spin down to spin up, probably
in 1988 (Rubin et al. 1997), which has also been confirmed by BeppoSAX
(Robba et al. 2001). The spectrum has an iron K fluorescence line at 6.4 keV, a
cyclotron absorption line near 20 keV and is well fitted with a powerlaw
and high energy cut-off at $\sim$ 16 keV (Clark et al. 1990,
Robba et al. 2001). X-ray eclipse phenomena were investigated by
Clark, Woo and Nagase (1994) with the GINGA data.

4U~1538--52, a wind fed HMXB pulsar with
a compact eccentric orbit (e $\sim$ 0.18) is very suitable
for the study of the orbital evolution and wind structure of the 
companion star. In this paper, we report pulse timing measurements with 
a new observation with the RXTE and also the out-of-eclipse spectral evolution 
with orbital phase. Better spectral coverage with BeppoSAX
(going down to 0.3 keV) enables us to constrain the column density values well. 
We consider the pulsar to be immersed in a spherically symmetric
stellar wind from the companion star and compare model column density
profiles with the observed profile.  

\section{Observations}

We observed the source with the Proportional Counter Array (PCA) of 
RXTE from 2003-07-31 to 2003-08-07 covering 
the out of eclipse phases for two binary orbits. There were twenty-five
observations, each corresponding to one orbit of the satellite. Each
observation was of 1.5--6.0 ks duration. For spectral analysis, we also used
an archival observation from RXTE spanning from 1997-01-01 to 1997-01-05,
having fourteen satellite orbit data.
Each observation had 1.6--10.0 ks of on-source time.
On an average, three PCUs (see next paragraph) were ON for the observations in 2003 
while all the five PCUs were ON for the observations made in 1997.
Both these datasets from RXTE provide a comprehensive orbital 
coverage. We have also used archival BeppoSAX data 
obtained between 1998-07-29 to 1998-08-01, covering one binary orbit.
There were forty segments in this observation; one for each satellite orbit,
with an exposure of 1.7--4.0 ks. The motivation for using the BeppoSAX 
observations is to have a better spectral coverage since the lower values of column 
densities are measured more accurately with BeppoSAX, which has a good response 
at low energies.

RXTE consists of a large-area proportional counter array (PCA), 
with five Xenon proportional counter units (PCUs) and sensitive
in the energy range of 2--60 keV with an effective area of 6250 cm$^{2}$
(Jahoda et al. 1996). It also consists of a high energy crystal
scintillation experiment (HEXTE; 15--200 keV; 1600 cm$^{2}$ area), and a
continuously scanning all-sky monitor (ASM; 2--10 keV; 90 cm$^{2}$).
The RXTE--ASM long term light curve of 4U~1538--52 folded at the orbital period
is shown in Figure 1 which clearly shows the eclipse with a slow ingress
and a sharp egress. The BeppoSAX observatory has a set of Narrow Field Instruments
(NFI) comprising one Low Energy Concentrator Spectrometer
(LECS; 0.1--10 keV, 22 cm$^{2}$ @ 0.28 keV), three Medium Energy
Concentrator Spectrometers (MECS; 1.3--10 keV, 150 cm$^{2}$ @ 6 keV),
one High Pressure Gas Scintillator Proportional Counter
(HPGSPC;  4--120 keV, 240 cm$^{2}$ @ 30 keV), and a Phoswich Detection System
(PDS; 15--300 keV, 600 cm$^{2}$ @ 80 keV).
A detailed description of BeppoSAX can be found in Boella et al. (1997).

\section{Timing Analysis \& Results}

Only the 2003 RXTE-PCA data were used for a timing analysis.
The light curve was extracted from event
mode data with time resolution of 0.125 s. Only photons detected in the top two 
layers were used. The photon arrival times were then corrected to the solar system barycenter.
Figure 2. shows the complete barycentered light curve for the August
2003 observation with RXTE-PCA.

The light curve shows variability at time scales other than the regular
526.85 s pulses from the neutron star. The pulsed emission is strong enough
for individual pulses to be seen in the energy range from 1 keV to 40 keV.
The pulse profile is double peaked with a main pulse and an inter pulse
which is about half the amplitude of the main pulse. The average pulse
period of the source was obtained by pulse folding and $\chi^{2}$ maximisation
method.

The light curve was then folded with the average spin period derived to get
pulse profiles at different orbital phases. Each pulse profile is an average
of two successive pulses binned into 256 pulse phase bins. A total of
fortysix such pulse profiles were generated, spanned over two binary orbits 
of 4U~1538--52. To get the arrival times of these pulses, we decomposed each
pulse profile into 128 Fourier components. The phase of the first Fourier
component which also has the highest amplitude was used to find the arrival
time delay of the respective pulses. Figure 3a shows the arrival time delay
curve.  

The arrival time delay curve when fitted with a circular orbital delay curve
(i.e. $e=0$) leaves out systematic residues as shown in Figure 3b indicating
an elliptical orbit. Subsequently, we fitted the pulse arrival time
delay with a function appropriate for elliptical orbit given below.
\begin{equation}
 \Delta t = \frac{a_x \sin i}{c} (1-e^2) \frac {\sin ( \nu + \omega )}{1+e \cos \nu} 
\end{equation}
\begin{equation}
\tan \frac{ \nu }{2} = \sqrt{ \frac {1+e}{1-e} } \tan \frac{E}{2}
\end{equation}
\begin{equation}
E-e \sin E = \frac {2 \pi}{P_{orb}}(t - T_0 - \frac{P_{orb}}{2 \pi} \omega) 
\end{equation}
\begin{equation}
T_0 = T_{\pi/2} - \frac{P_{orb}}{4} 
\end{equation}
It was seen that the errors in the individual measurements of pulse arrival
times were greater than the error due to photon counting statistics, possibly
due to some aperiodic intensity variation over a range of time scales.
While fitting the arrival delay curve to an elliptical binary orbit,
the statistical uncertainties on the arrival time measurements were
multiplied by a constant factor to obtain a reduced $\chi ^2$ value of 1.0.
This method was employed so that one can compare the orbital parameter
estimations with the earlier measurements (Clark 2000).
For a circular orbit, the rescaled uncertainties gives a best fit
with a $\chi ^2$ value of 64 for 43 degrees of freedom. Comparing
this with the reduced $\chi ^2$ value of 1 for an elliptical orbit
as described above, we reconfirm an elliptical orbit for this system.
The value of the five free parameters
for an elliptical fit, namely $P_{spin}$, $a_x\sin i$, e, $\omega$, and
$T_{\pi/2}$ are listed in Table 1. We combined the new epoch measured from the RXTE data
taken in 2003 with the earlier measurements listed in Clark (2000). The
mid-eclipse times derived from light curves obtained using the UHURU and ARIAL V 
were not included because the X-ray eclipse is asymmetric as seen in the
folded ASM light curve of Figure 1. As the X-ray source goes into eclipse
gradually and comes out of the eclipse rather sharply, the mid-eclipse
time determined from the light curve only is likely to be inaccurate and
have some systematic errors. Fitting the remaining 9 mid-eclipse times
with a linear function gives a reduced $\chi^2$ of 1.3 for 7 degrees of
freedom. The residuals from a linear fit to the mid-eclipse times are
shown in Figure 4. A quadratic fit improves the reduced $\chi^2$ by a mere 0.01
(for 6 degrees of freedom). If the mid-eclipse
time record is fitted with an orbital evolution corresponding to the value
reported in Clark (2000), a reduced $\chi^2$ of 2.5 is obtained, which is
larger than the same for a linear fit. We, therefore, conclude that there
is no evidence for an orbital evolution in 4U~1538--52 from the
available data.  

\section{Spectral Analysis \& Results}

For the RXTE data, we took the Standard 2 data products of PCA and extracted 
the source spectra for each dataset using the tool saextrct v 4.2d for the 
appropriate good time intervals. Background data files were generated with 
the tool runpcabackest using background models appropriate for the source
brightness and the epoch of the RXTE observation, as provided by the
PCA calibration team. The background subtracted source spectra
were analyzed with the spectral analysis package XSPEC v 11.2.0
(Shafer, Haberl $\&$ Arnaud 1989).
Model spectra were convolved with the detector response matrix and
fitted to the observed pulse phase averaged spectrum
in the spectral range of 3--20 keV and the important spectral parameters
were determined. The spectral model used 
was a powerlaw alongwith a high energy cut off and a Gaussian line with centre 
energy $\sim$ 6.4 keV with a photoelectric absorption 
using Morrison and McCammon cross-sections (Morrison and McCammon 1983).
All the parameters were kept free in the 2003 dataset while for the
1997 dataset, the line centre energy and the width were kept frozen at
6.4 keV and 0.01 keV respectively.

The source and background spectra of BeppoSAX were extracted from the MECS and
LECS detectors using circular regions of radius 4$'$ and 
8$'$ respectively. The energy range chosen for MECS
was (1.8--10.0) keV while that for LECS was (0.3--4.5) keV, 
in which the respective instruments have large effective area and the
spectral responses are well understood. For all the three observations,
the MECS and LECS phase averaged out-of-eclipse spectra were fitted
simultaneously with the relative normalization of the two instruments allowed
to vary. All the spectra were suitably binned to allow use of $\chi^{2}$
statistics. Both the MECS and LECS data of BeppoSAX were simultaneously 
fitted using XSPEC v 11.2.0. Due to the 
restricted energy range (1.8--10.0 keV), no high energy cutoff was required. 
The out of eclipse pulse phase averaged spectrum was first fitted with a
simple model comprising of a powerlaw and a Gaussian line alongwith the photoelectric absorption using 
Morrison and McCammon cross-sections. The values of the line centre energy and 
line width as obtained in this fit were then kept frozen to fit spectra 
at different orbital phases.

The RXTE-PCA and the BeppoSAX spectra were all well fitted with 
a relatively simple model. Robba et al. (2001) had detected 
marginal evidence of a soft excess at $\sim$ 1 keV which was 
resolved by modeling the spectrum with one additional component,
either as a black-body of temperature 0.08 keV or a bremsstrahlung of 
temperature 0.1 keV. We also found that the inclusion of a soft
black-body component in the phase averaged out-of-eclipse spectrum
from BeppoSAX marginally improves the fit. However, for orbital
phase resolved spectroscopy, we have not included a soft component as
it lies outside the range of the RXTE spectrum and in the orbital
phase resolved spectrum with BeppoSAX, it will be too faint to constrain.
In total, there were seventy nine spectra to be fitted and with
the above mentioned model, they gave reasonable reduced $\chi^{2}$ values
in the range of 0.8 to 1.5 for RXTE (with 27 dof), and
0.6 to 1.3 for BeppoSAX (with 63 dof).
Representative spectra, one each from the RXTE-PCA and BeppoSAX are 
shown in Figures 5 \& 6 respectively.
The RXTE-PCA spectra analysed here also does not 
include the energy range of the cyclotron absorption feature.

The spectral parameters obtained from the fitting of 
the three datasets used in our analysis were investigated 
for their variation against the orbital phase of the 
pulsar. The photon index of the power law is seen to remain more
or less constant over the binary phase having 
values between 1.0 to 1.5. The fluorescent iron-line 
flux measured with the average spectrum taken over 2000-3000 s,
also does not show any considerable variation along
the orbit. The cut-off energy measured with RXTE-PCA has a value $\sim$
14 keV and the e-folding energy is $\sim$ 7 keV. These are slightly
different from the values of the cut-off energy ($\sim$ 16 keV) and
e-folding energy ($\sim$ 10 keV) measured with BeppoSAX in a broad
energy band (Robba et al. 2001). We detected a notable variation in
the equivalent hydrogen column density with orbital phase as shown in
Figure 7. It shows a smooth variation over orbital phase, increasing
gradually by an order of magnitude as the pulsar 
approaches eclipse (mid-eclipse is defined by phase zero). 
At orbital phases far from the eclipse, the
column density has a value of $\sim$ 1.5 $\times$ 10$^{22}$ H atoms cm$^{-2}$.
In figures 1, 5, 6, \& 7, the orbital phase is measured with respect 
to the mid-eclipse time.

\section{Discussion}

A pulse period of 526.85 s of the pulsar 4U~1538--52 determined from
RXTE-PCA observation in 2003 shows that the spin-up trend of 4U~1538--52
first detected with BATSE during the 1990s has continued since then. The
overall spin-up time scale ($ \dot P/P$) during the five years of BATSE
observation reported by Rubin et al. (1997) was
9.6 $\times$ 10$^{-12}$ s$^{-1}$. The average spin-up rate between the
end of BATSE observations and the later observations with RXTE (1997,
Clark 2000), Beppo-SAX (1998, Robba et al. 2001) and again with RXTE
(2003, present work) is 1.4, 1.2 and 7.4 $\times$ 10$^{-11}$ s$^{-1}$,
significantly larger than the average spin-up rate during the BATSE era. An
increase in spin-up rate is expected if the overall X-ray luminosity had
increased after the BATSE observations. However, it is very
impractical to compare the pulsed hard X-ray flux measured by BATSE and the
total X-ray flux measurements with the later instruments in a different
energy band.

The value of  $a_x \sin i$ that we obtained from pulse timing of the RXTE-PCA
data in 2003 is 53.1 $\pm$ 1.5 lt-s, which is consistent with the BATSE result
and slightly smaller than that obtained from the 1997 RXTE data. We note here
that the accuracy of orbital parameter determination in slow pulsars like 4U~1538--52 is
considerably poorer than in fast pulsars like SMC X-1, Her X-1 etc.
In analysis of the BATSE data, Rubin et al. (1997) considered a circular
orbit whereas analysing pulse arrival times from two complete binary orbits
we found the orbit of 4U~1538--52 to be eccentric with e = 0.18 $\pm$ 0.01.
Our result is consistent with the results obtained from a 1997 observation of
this source with the RXTE-PCA (Clark 2000).
The value of $\omega$ determined from the 2003 observation is smaller
compared to that found by Clark from the 1997 observation. If the difference
is due to apsidal motion then the rate of apsidal motion is
$\dot \omega = -3.8^o \pm 2^o.2$ yr$^{-1}$. 
We found no significant evidence for orbital evolution in the system as
reported from earlier measurements (Clark 2000). 
The value of $\dot P_{orb}/P_{orb}$ determined from all the available
mid-eclipse times is $(2.5 \pm 1.9) \times 10^{-6}$ yr$^{-1}$.
Combining the pulse timing results on 4U~1538--52 obtained from our
observation with the earlier reported measurements, we are able to
rule out an orbital decay in this system for which Clark (2000) found
marginal evidence. The new upper limit presented here on possible changes
of the orbital period of 4U~1538--52 supersedes previous limits considerably.
\footnote{After this paper was submitted we came to know about a
paper by Baykal, Inam and Beklen (2006). These authors have used a different
analysis method and reached to similar conclusion about the orbital
period evolution of 4U 1538--52}

Tidal interaction between the two stars and mass loss from the companion
star are the two main effects that cause orbital decay in the HMXB systems.
The orbital evolution of several HMXB pulsar systems (Cen X-3, SMC X-1 and
LMC X-4) is very well documented. The companion star mass, and the size of
the binary orbit in 4U~1538--52 is very similar to the above mentioned
binaries, and therefore, the tidal effect that is expected in 4U~1538--52
should be as strong as in the three other HMXBs. In the present work, we
find no clear evidence of an orbital decay in 4U~1538--52. If at all, the
system shows a positive orbital period derivative, like the enigmatic system
Cyg X-3 (Singh et al.  2002). This indicates that mass loss may play a
prominent role in the evolution of this system.
Tidal interaction between the neutron star and the companion in close HMXBs
should also circularise the orbit. The orbits of most of the close (orbital
period less than 4 days) HMXBs are highly circularised with eccentricity
less than 0.006. But the orbit of 4U~1538--52 is not yet circularised and
shows no orbital decay. Therefore, it is likely that 4U~1538--52 is 
a relatively young system compared to the other HMXBs. Using the mass, radius 
and luminosity of the companion star QV Nor from the optical observation 
(Reynolds et al. 1992) the approximate time for tidal circularisation of this
system works out to be $2 \times 10^{3}$ yrs (Lecar, Wheeler \& McKee,1976).
Since the system is still eccentric, it is possible that the age of the system
after the supernova explosion is of the order of a few thousand years; very
small compared to the lifetime of HMXBs.  

The eccentricity derived using the two RXTE observations in 1997 (Clark 2000)
and 2003 (reported here) is likely to be correct for two reasons. The RXTE-PCA
has a large effective area compared to the earlier missions and in the
energy range of PCA more photons are detected compared to CGRO-BATSE 
using which Rubin et al (1997) did not detect any eccentricity. If the
eccentricity is high, then either the binary system is young or the
circularisation time is actually much longer than a few thousand years
estimated using Lecar's method (1976). On the other hand, a binary is
unlikely to become very X-ray active by wind accretion within just a few
thousand years of the supernova explosion. It is possible that the
circularisation mechanism works very efficiently in systems with Roche-lobe
overflow like LMC X-4 while accretion mechanism in 4U 1538--52 is likely
to be driven by stellar wind.

Lack of any substantial variation in the photon index, the cut-off energy,
and the e-folding energy over the binary phase of 4U~1538--52 suggests that
the continuum X-ray spectrum of the pulsar is hardly affected during its
revolution along the orbit. This indicates a nearly constant accretion rate
and identical conditions in the accretion column throughout the binary orbit.
In the HMXB pulsar GX 301-2, orbital phase resolved spectroscopy with RXTE-PCA
showed that the spectral parameters remain almost constant throughout the
orbit though the overall accretion rate and wind density, and clumpiness of
the wind changes considerably with the binary phase (Mukherjee and Paul 2004).
On the other hand, in 4U~1538--52 the equivalent hydrogen column density
shows a smooth variation over the orbital phase. We compare the observed
column density profile with a model estimated by assuming a spherically
symmetric Castor, Abbott \& Klein (CAK 1975) 
wind from the companion star. The velocity profile of the wind is:
\begin{equation}
   v_{\mathrm wind} = {v_{\infty}}\sqrt{1-\frac {{R_\mathrm c}}{r}}\,
\end{equation}
where $v_{\infty}$ is the terminal velocity for the
stellar wind, R$_{\mathrm c}$ is the radius of the companion
and r is the radial distance from centre of the companion star.
The column density profiles were derived using a numerical integration
along the line of sight from the pulsar to the observer for three
different inclination angles 65$^\circ$,
75$^\circ$ and 85$^\circ$ respectively. In Figure 7 we show the 
equivalent hydrogen column density measured with BeppoSAX and RXTE-PCA
as a function of the orbital phase with respect to the mid-eclipse time.
A mass-loss rate of $\sim$10$^{-6}$ M$_\odot$ yr$^{-1}$ and 
v$_\infty$ $\sim$1000 km s$^{-1}$ were assumed.
The model calculations of the absorption column density for different
inclination angles when superposed on the observed values indicate
reasonable agreement as can be seen in Figure 7. This indicates that a spherically symmetric stellar 
wind from the companion star may produce the observed orbital dependence of 
the column density for certain range of the orbital inclination.

We note here that the we have not done any fitting of the column density
measurements and the model column density values at different orbital
phases (Figure 7). For the wind density model used here, the phase
resolved column density values depend on the mass loss rate, the
terminal velocity and the inclination angle of the binary orbit. The
data presented here is not suitable for such a detailed analysis but
only shows that such a model is consistent with the observations.
In eclipsing X-ray binaries like 4U 1538--52, the eclipse duration
puts strong constraints on the orbital inclination. We point out that
the orbital phase resolved column density measurements can be an
independent way of estimating the orbital inclination, especially for
non-eclipsing X-ray binaries. We also note that in several eclipsing
and wind-fed HMXBs, the eclipse ingress is more gradual than the
eclipse egress (Haberl et al. 1989, Feldmeier et al. 1996, Naik \& Paul 2004). 
From the RXTE-ASM light curves of many eclipsing HMXB pulsars we
have verified that this effect is more pronounced at lower energies.
This ingress-egress asymmetry, which varies in degree from source to
source is partly due to absorption by accretion material trailing
the X-ray source in its orbit. For example, Vela X-1 (Feldmeier et al. 1996) 
has a much larger ingress-egress asymmetry and LMC X-4 (Naik \& Paul 2004)
has a much smallter asymmetry compared to 4U 1538--52 (Figure 1).
We think that the slight asymmetric accretion column distribution in
the case of 4U 1538-52 on two sides of the eclipse (Figure 7) can also
partly be due to some trailing accreting material.

In several other HMXBs like 4U~1700--37 (Haberl et al. 1989),
4U~1907+09 (Roberts et al. 2001) and GX 301--2 (Leahy 1991,
Pravdo \& Ghosh 2001, with RXTE), it has been shown that a simple
spherical wind emanating from the companion star is not sufficient to
explain the column density profile. Density enhancement due to a trailing 
stream or a disk from the companion arising due to the dynamical 
effects of the pulsar was invoked to explain the said profile.
In the case of GX 301--2, detailed observations with RXTE-PCA revealed
that neither a disk nor a stream was sufficient to explain the observed
profile (Mukherjee and Paul 2004). Considering the results for these
HMXBs, the present work seems to carry an interesting implication.
Clark, Woo and Nagase (1994) have studied the column density variation
during eclipse egress for this particular source with GINGA data.
Two HMXB pulsars in which similar increase in absorption column
density pattern was seen near eclipse are X1908+075 (Levine et al. 2004) and
SMC X-1 (Woo et al. 1995) which also may have isotropic wind pattern
from the companion stars. In the present study we investigated the 
out-of-eclipse variation of the column density and found that a simple model
with a spherically symmetric stellar wind describes the observations well.
Since 4U~1538--52 has a moderate X-ray luminosity ($\sim$ 10$^{36}$
erg s$^{-1}$), it is not expected to cause any significant perturbation in
the acceleration of the stellar wind through X-ray ionization as compared to
high luminosity sources like Cen X-3 (Day \& Stevens 1993) and SMC X-1
(Woo et al. 1995) with luminosities of $\sim$ 10$^{38}$ erg s$^{-1}$ in
the high states. Observation of 4U~1538--52 covering the
eclipse ingress and egress with high throughput telescope like XMM-Newton
and SUZAKU which also have good low energy response will allow one to carry
out a detailed study of the stellar wind structure in this source and also
determine the extent to which the local X-ray photoionisation affects the
stellar wind acceleration.

\section*{Acknowledgements}
We thank an anonymous referee for some useful suggestions that helped us
to improve the paper.
This research has made use of data obtained from the High Energy Astrophysics 
Science Archive Research Center (HEASARC), provided by NASA's Goddard Space 
Flight Center. We also thank the BeppoSAX team for making 
the data available. UM and HR would like to
acknowledge the Kanwal Rekhi Scholarship of TIFR Endowment Fund for partial
financial support.

\section*{References}

Boella et al., 1997, {\it A\&AS}, {\bf 122}, 299 \\
Castor, J.~I., Abbott, D.~C., \& Klein, R.~I., 1975, {\it ApJ}, {\bf 195}, 157 \\
Clark, G.~W., Woo, J.~W., Nagase, F., Makishima, K., \& Sakao, T., 1990, {\it ApJ}, {\bf 353}, 274 \\
Clark, G.~W., Woo, J.~W., \& Nagase, F., 1994, {\it ApJ}, {\bf 422}, 336 \\
Clark, G.~W., 2000, {\it ApJ}, {\bf 542}, 131L \\
Davison, P.~J.~N., Watson, M.~G., \& Pye, J.~P.,  1977, {\it MNRAS}, {\bf 181}, 73
Day, C.~S.~R., \& Stevens, I.~R., 1993, {\it ApJ}, {\bf 403}, 322 \\
Feldmeier, A., Anzer, U., Boerner, G., \& Nagase, F., 1996, {\it A\&A}, {\bf 311}, 793 \\
Giacconi et al., 1974, {\it ApJS}, {\bf 27}, 37 \\
Haberl, F., White, N.~E., \& Kallman, T.~R., 1989, {\it ApJ}, {\bf 343}, 409 \\
Jahoda et al., 1996,
    In: Siegmund O.H.W., Gummin M.A. (eds.) EUV, X-Ray and Gamma-Ray Instrumentation for Astronomy VII. SPIE 2808, p. 59
Leahy, D.~A., 1991, {\it MNRAS}, {\bf 250}, 310L \\
Lecar, M., Wheeler, J.~C., \& McKee, C.~F., 1976, {\it ApJ}, {\bf 205}, 556L \\
Levine, A.~M., Rappaport, S.~A., Zojcheski, G., 2000, {\it ApJ}, {\bf 541}, 194L \\
Levine, A.~M., Rappaport, S., Remillard, R., \& Savcheva, A., 2004, {\it ApJ}, {\bf 617}, 1284 \\
Makishima, K., Koyama, K., Hayakawa, S., \& Nagase, F., 1987, {\it ApJ}, {\bf 314}, 619 \\
Morrison \& McCammon, 1983, {\it ApJ}, {\bf 270}, 119 \\
Mukherjee, U., \& Paul, B., 2004, {\it A\&A}, {\bf 427}, 567 \\
Nagase, F., Corbet, R.~H.~D., Day, C.~S.~R., Inoue, H., Takeshima, T., Yoshida, K., \& Mihara, T., 1992, {\it ApJ}, {\bf 396}, 147 \\
Naik, S., \& Paul, B., 2004, {\it ApJ}, {\bf 600}, 351 \\
Parkes, G.~E., Murdin, P.~G., \& Mason, K.~O., 1978, {\it MNRAS}, {\bf 184}, 73 \\
Pravdo, S.~H., \& Ghosh, P., 2001, {\it ApJ}, {\bf 554}, 383 \\
Reynolds, S.~A., Bell, S.~A., \& Hilditch, R.~W., 1992, {\it MNRAS}, {\bf 256}, 631 \\
Robba, N.~R., Cusumano, G., Orlandini, M., dal Fiume, D., \& Frontera, F., 1992, {\it ApJ}, {\bf 401}, 685 \\
Robba, N.~R., Burderi, L., Di Salvo, T., Iaria, R., \& Cusumano, G., 2001, {\it ApJ}, {\bf 562}, 950 \\
Roberts et al., 2001, {\it ApJ}, {\bf 555}, 967 \\
Rubin, B.~C., Finger, M.~H., Scott, D.~M., \& Wilson, R.~B., 1997, {\it ApJ}, {\bf 488}, 413 \\
Shafer, R.~A., Haberl, F., \& Arnaud, K.~A., 1989, XSPEC: An X-ray Spectral Fitting Package, ESA TM-09 (Paris:ESA)\\
Singh, N.~S., Naik, S., Paul, B., Agrawal, P.~C., Rao, A.~R., \& Singh, K.~Y., 2002, {\it A\&A}, {\bf 392}, 161 \\
Wojdowski, P., Clark, G.~W., Levine, A.~M., Woo, J.~W., \& Zhang, S.~N, 1998, {\it ApJ}, {\bf 502}, 253 \\
Woo, J.~W., Clark, G.~W., Blondin, J.~M., Kallman, T.~R., \& Nagase, F., 1995, {\it ApJ}, {\bf 445}, 896 \\

\clearpage

\begin{figure}
\vskip 10. cm
\includegraphics{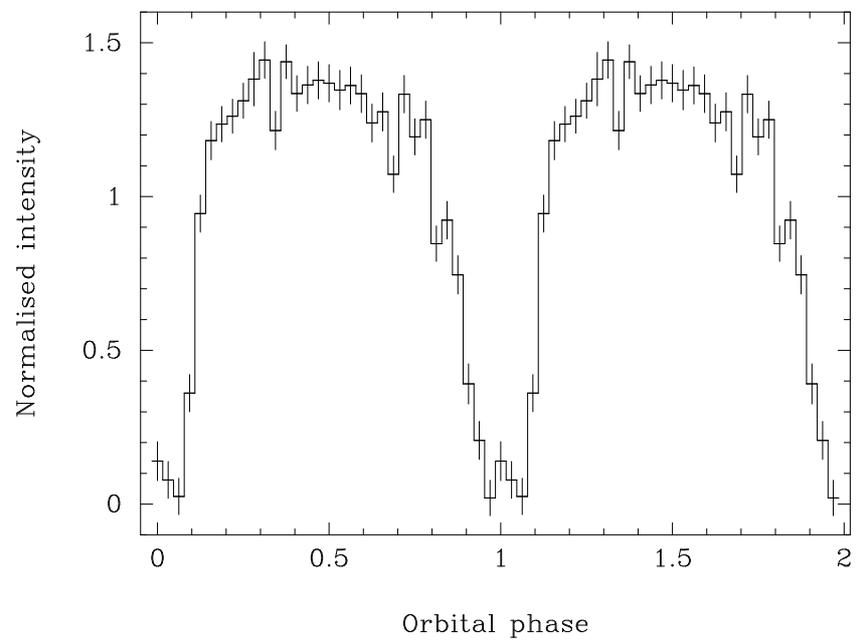}
\caption{The RXTE-ASM light curve folded at the orbital period of
$\sim$ 3.73 days. Two orbital cycles are shown for clarity.}
\end{figure}

\begin{figure}
\vskip 10. cm
\includegraphics{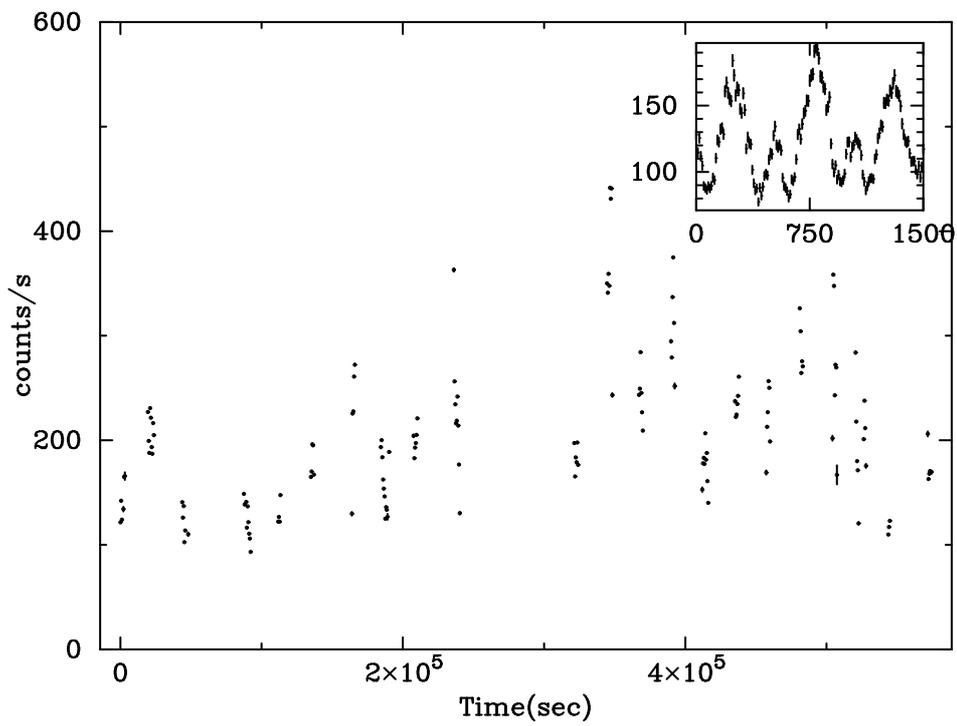}
\caption{This plot shows the complete barycentered light curve for August
2003 data with a time bin of 526.85 s, the same as the pulse period. The inset
is an expanded view of a part of the total light curve with a time bin
size of 10 s. Individual double peaked pulses can be clearly seen.}
\end{figure}

\begin{figure}
\vskip 10. cm
\includegraphics{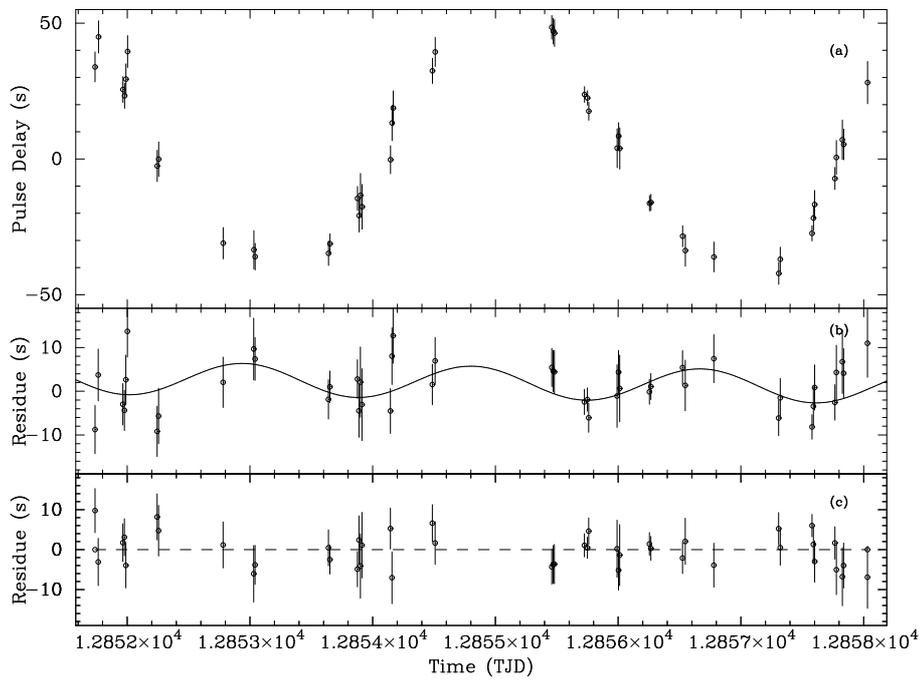}
\caption{(a) Pulse arrival time delays measured from the RXTE-PCA observation
in 2003, (b) residue of a circular fit to pulse arrival times, and
(c) residue of an elliptical fit to pulse arrival times are shown here.}
\end{figure}

\begin{figure}
\vskip 10. cm
\includegraphics{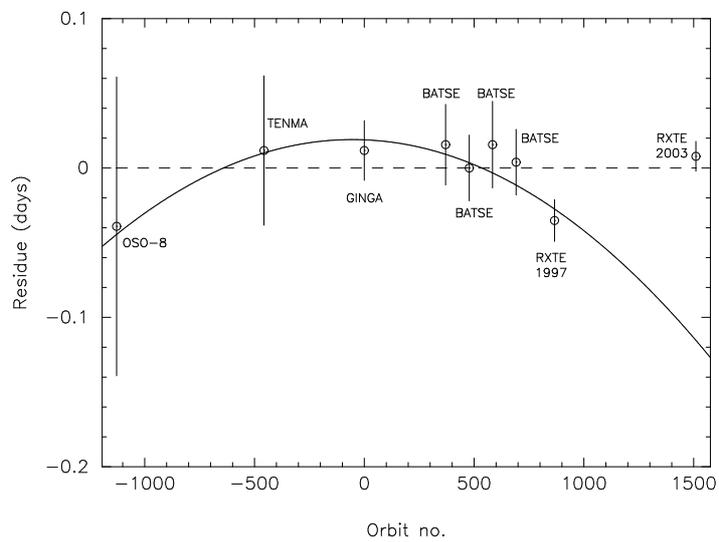}
\caption{Differences between measured mid-eclipse times and a linear fit to
the same are shown here. Solid line indicates the quadratic fit of mid-eclipse
times till the 1997 observation (Clark 2000).}
\end{figure}

\begin{figure}
\vskip 10. cm
\includegraphics{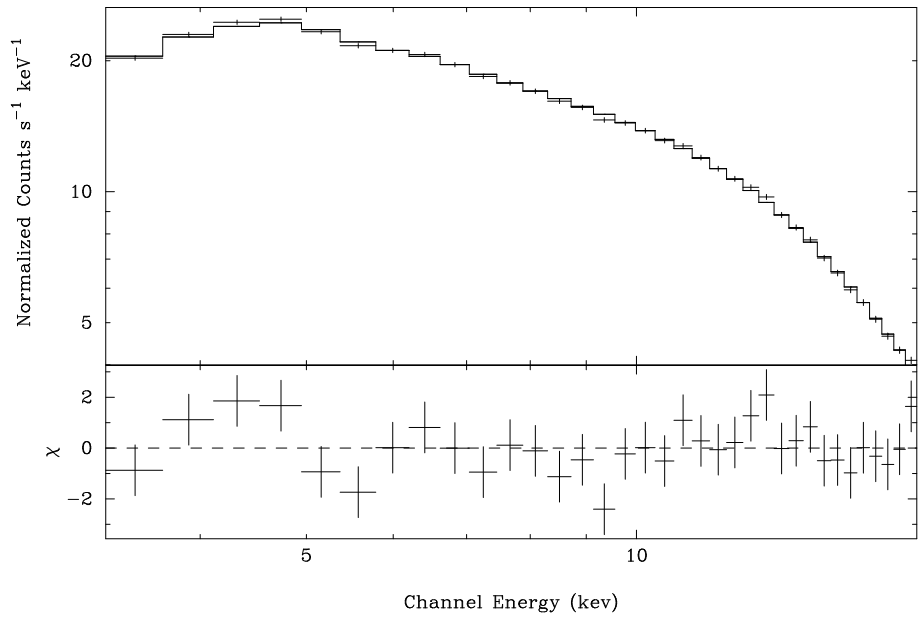}
\caption{RXTE-PCA spectrum from an observation in 2003 corresponding
to the orbital phase $\sim$ 0.28. The residuals are shown in the lower panel.
Soft excess was not included as it lies outside the energy range.}
\end{figure}
\clearpage

\begin{figure}
\vskip 10. cm
\includegraphics{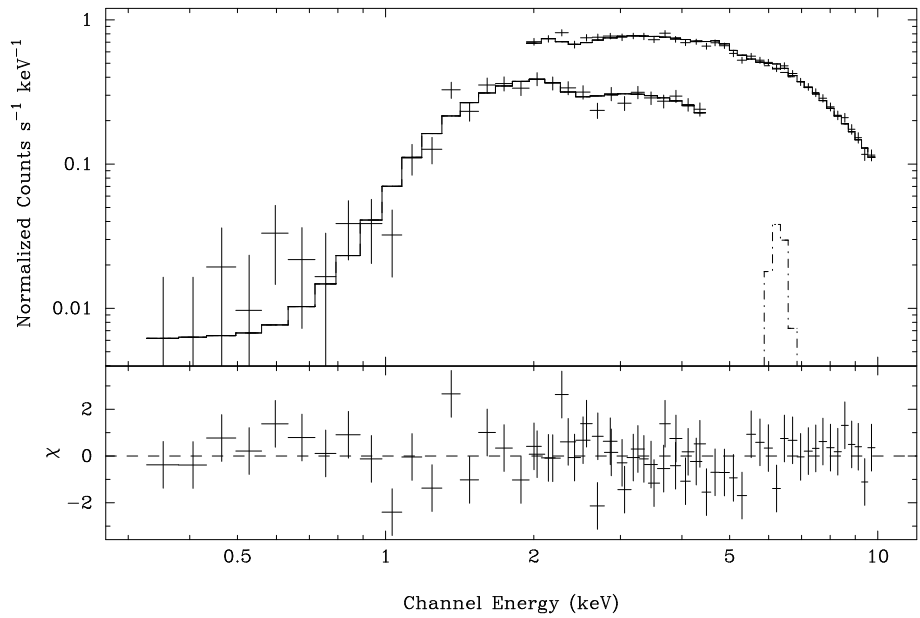}
\caption{A BeppoSAX spectrum corresponding to orbital phase $\sim$ 0.59
with the residuals in the lower panel. The narrow Gaussian component
at 6.4 keV shown here is the iron fluorescence line.}
\end{figure}
\clearpage

\begin{figure}
\vskip 10. cm
\includegraphics{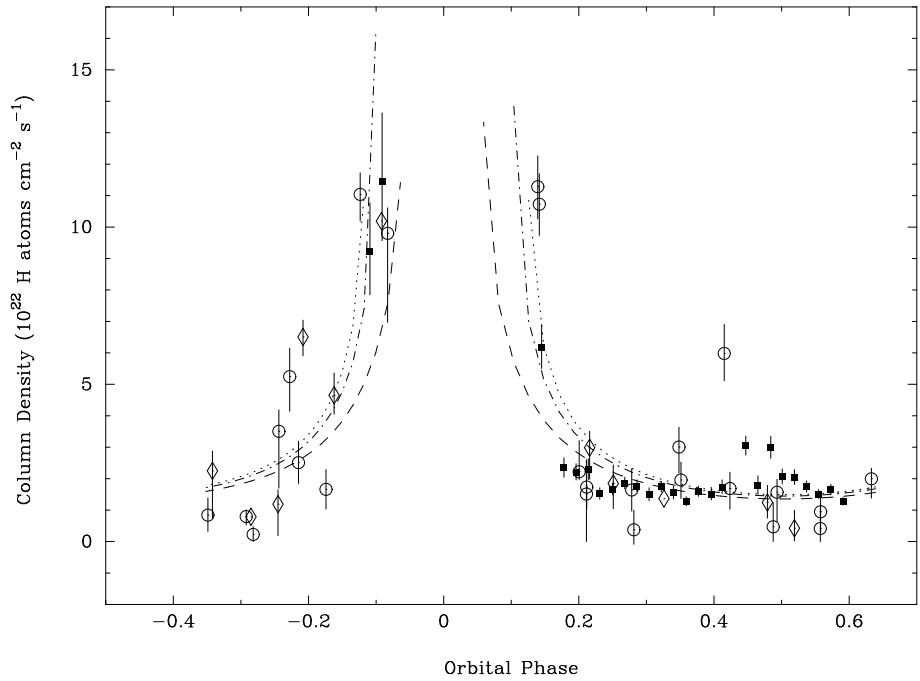}
\caption{Variation of column density versus orbital phase with respect to
the mid-eclipse time is shown here. The
dashed line represents the model for inclination angle of 65$^\circ$, the
dashed-dotted line for inclination angle of 75$^\circ$ and the dotted line for
inclination angle 85$^\circ$. Diamonds, filled squares, and circles denote
measurements from observations made with RXTE in 1997, BepposSAX in 1998,
and RXTE in 2003 respectively. The error-bars shown correspond to a 90$\%$ confidence interval.}
\end{figure}
\clearpage

\vskip 0.5cm
\begin{table}
\caption{The spin and orbital parameters of 4U~1538--52 with
$1\sigma$ errors.}
\vskip 0.5cm
\begin{tabular}{ll}
\hline
\\
Parameter&Value\\
\\
\hline
\\
$P_{spin}$ & $526.849 \pm 0.003$ s \\
$a_x \sin i$ & $53.1 \pm 1.5$ lt-s \\
e & $0.18 \pm 0.01$ \\
$\omega$ & $40^o \pm 12$ \\
$T_{\pi/2}$ & $52851.33 \pm 0.01$ MJD \\
$a_o$ & $47221.463 \pm 0.012$ lt-s\\
$P_{orb}$ & $3.728382 \pm 0.000011$ d \\
\\
\hline\\
\small{$T_{\pi/2}(N) = a_0 + P_{orb}N$}
\end{tabular}
\end{table}

\end{document}